\documentclass[global,twocolumn]{svjour}
\usepackage{graphics}
\usepackage{amssymb}
\usepackage{dsfont}
\journalname{Applied Physics A}
\begin{document}
\title{Quantum point contact due to Fermi-level pinning and doping profiles in semiconductor nanocolumns}

\author{K. M. Indlekofer
\and M. Goryll
\and J. Wensorra
\and M. I. Lepsa
}                     
\offprints{m.indlekofer@fz-juelich.de}          
\institute{
Center of Nanoelectronic Systems for Information Technology, IBN-1,
Research Centre J\"ulich, D-52425 J\"ulich, Germany
}
\date{Received: date / Revised version: date}
%
\maketitle
\begin{abstract}
We show that nanoscale doping profiles inside a nanocolumn
in combination with Fermi-level pinning at the surface
give rise to the formation of a saddle-point in the potential profile.
Consequently, the lateral confinement inside the channel varies along the transport direction,
yielding an embedded quantum point contact.
An analytical estimation of the quantization energies will be given.
\end{abstract}
\section{Introduction}
\label{sec:1}
Structures based on quasi-1D nanocolumns are not only of major interest for basic research
in the area of quantum transport but can also be considered as prototype systems of future
nanodevices for information technology. Research in this field has attracted much attention
recently due to the fact that nanocolumns can be fabricated either by employing self-organized
growth modes or by lithographic patterning using a top-down approach
\cite{Ref_NL5_2470,Ref_APL81_4458,Ref_PhysE25_2004,Ref_NL3_149,Ref_NL5_981}.
In any case, due to their lateral dimensions in the nanometer regime, quantum confinement effects are
expected to arise in such nanocolumns, leaving their signatures in the electronic transport properties.
As a common feature, most of the application-relevant semiconductor material systems exhibit Fermi-level pinning
at the outer surface due to surface states. Depending on the energetic position of the neutrality point,
either surface depletion or surface accumulation occurs.
In the following, we will consider the class of surface-depleted semiconductors such as GaAs or GaN.
For nanocolumn systems, the spatial extent of the system can approach the size of the depletion region
and lateral quantum confinement effects gain considerable influence on the electronic spectrum.
The depletion length strongly depends on space charge, for example due to ionized doping impurities.
Hence, doping profiles in combination with surface pinning have a significant influence on the
potential profile on a nanometer length scale inside the nanocolumn channel.

We will discuss how nanoscale doping profiles inside a nanocolumn
in combination with Fermi-level pinning at the surface
give rise to the formation of a saddle-point in the potential.
As a result, the lateral confinement inside the channel varies along the transport direction,
yielding an embedded quantum point contact without the need for any heterostructure.
In the following sections, we give an analytical estimation of the quantization energies
and discuss a realistic example.

\section{Nanocolumn confinement potential}
\label{sec:2}
\begin{figure}
\resizebox{0.45\textwidth}{!}{
  \includegraphics{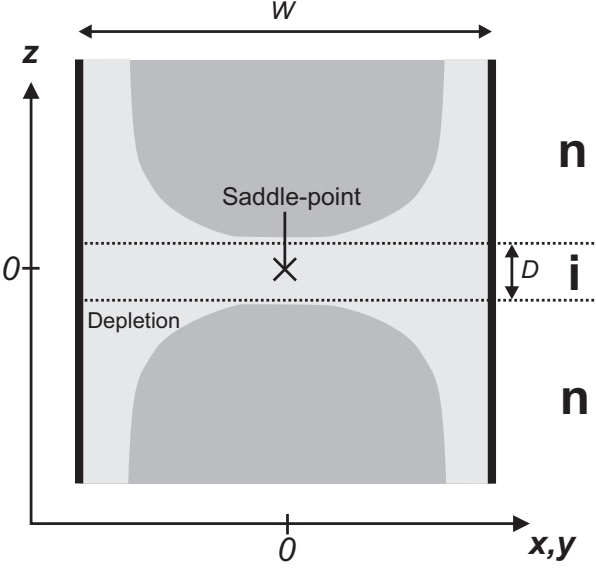}
}
\caption{
Cross-sectional view of a nanocolumn of diameter $W$ and embedded $n-i-n$ doping sequence.
Depleted regions is visualized in a light-gray color.
Due to Fermi-level pinning at the surface, a saddle-point in the potential is formed.
}
\label{fig:column}
\end{figure}

In order to visualize the interplay between surface pinning and doping profiles,
we now consider a typical example of a cylindrical nanocolumn with an embedded
$n-i-n$ doping profile as depicted in Fig.~\ref{fig:column}.
Here, the main feature consists of a nanoscale intrinsic ($i$) layer
between two $n$-doped contact regions, perpendicular to the transport direction.

The Poisson equation for the conduction band profile (electronic potential energy) $E_C$ reads as
\begin{equation}
\label{eq:poisson}
\Delta E_C(\vec{x})=\frac{e}{\epsilon_0\epsilon_r}\rho(\vec{x}),
\end{equation}
for a given charge density $\rho$,
and $\Delta\equiv\frac{\partial^2}{\partial x^2}
+\frac{\partial^2}{\partial y^2}
+\frac{\partial^2}{\partial z^2}$.
As a boundary condition, we assume Fermi-level-pinning at the outer surface of the nanocolumn.
Furthermore, the system is assumed to be rotational invariant within the $x,y$-plane.
Note that $E_C$ and $\rho$ have to be determined self-consistently,
containing contributions from free electrons and ionized donors.
In general, such a nanoscale problem requires a quantum statistical approach and cannot be solved
analytically. In the following, we thus provide a simplified analytical estimation for the expected behavior.

\begin{figure}
\resizebox{0.45\textwidth}{!}{
  \includegraphics{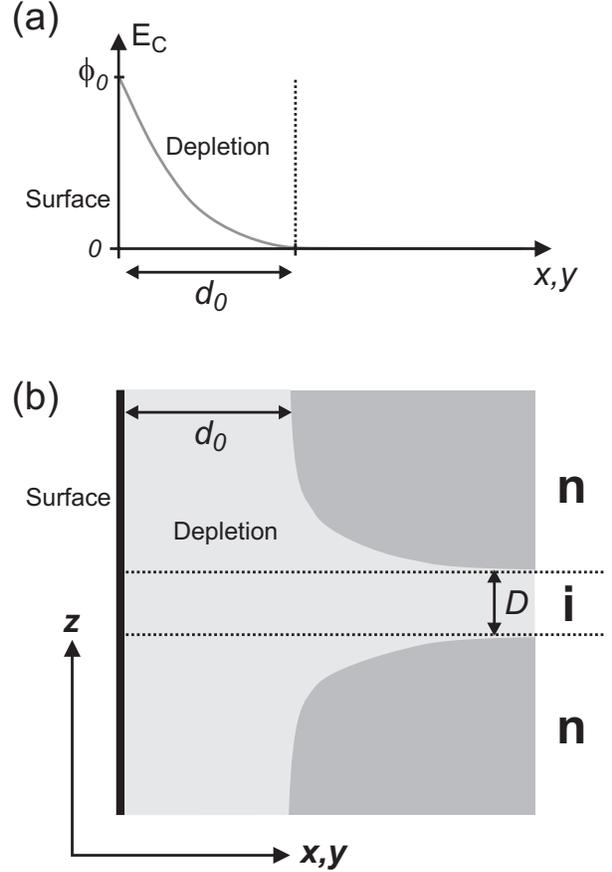}
}
\caption{
(a) Schottky depletion for the homogeneous case.
$d_0$ and $\phi_0$ denote the depletion length and the surface pinning, respectively.
(b) $n-i-n$ structure close to the surface.
The depleted region is visualized in a light-gray color.
At the outer ends ($z\to\pm\infty$), the homogeneous depletion length $d_0$ is retained.
}
\label{fig:nin}
\end{figure}

For a homogeneously doped nanocolumn, the potential profile can be constructed to a good approximation
by the well-known harmonic Schottky depletion model as depicted in Fig.~\ref{fig:nin}(a).
For a given donor concentration $N_D$ one obtains a depletion length of
\begin{equation}
\label{eq:d0}
d_0=\sqrt{\frac{2\phi_0\epsilon_0\epsilon_r}{e^2N_D}}.
\end{equation}
This expression for $d_0$ represents a good approximation for the cylindrical column if $d_0<W/2$.
(For a more general discussion of the Poisson problem in systems with cylindrical symmetry
see Refs.~\cite{Ref_SSE46_885,Ref_JE_63_1115}].)

\begin{figure}
\resizebox{0.45\textwidth}{!}{
  \includegraphics{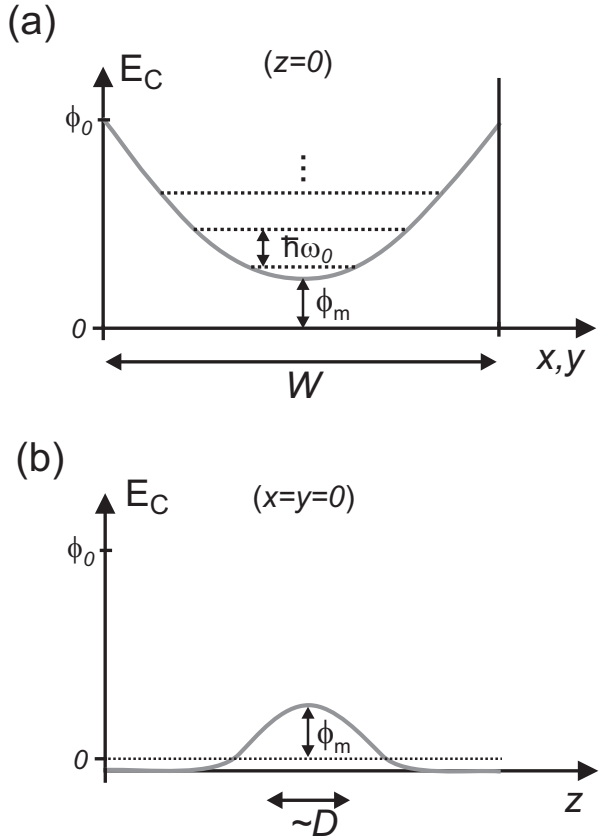}
}
\caption{
(a) Potential energy within the $x,y$-plane at $z=0$.
$\phi_0$, $\phi_m$, and $\hbar\omega_0$ denote the pinning level, potential minimum,
and confinement quantization energy,
respectively. The lowest quantized levels are depicted as dotted lines.
(b) Potential profile along the center axis (i.e., $x=y=0$).
The width of the barrier is determined by the length scale $D$.
The Fermi-energy $E_F\equiv 0$ is chosen as the energy reference.
}
\label{fig:harmpot}
\end{figure}

For the $n-i-n$ structure, we insert an intrinsic layer of thickness $D$
as sketched in Fig.~\ref{fig:nin}(b).
Due to $\Delta E_C\approx 0$ within the intrinsic layer,
the resulting $E_C$ must exhibit compensating curvatures in $x,y$ and $z$.
The overall potential within the intrinsic layer thus must exhibit a saddle-point
as visualized in Fig.~\ref{fig:column},
exhibiting a minimum (confinement in the constriction) in $x,y$-direction
and a maximum (barrier) in $z$-direction.
The cross-sectional potential within the $x,y$-plane at $z=0$ is plotted
qualitatively in Fig.~\ref{fig:harmpot}(a), whereas the potential
along the center axis ($x=y=0$) is sketched in Fig.~\ref{fig:harmpot}(b).
The resulting lateral quantization energy $\hbar\omega_0$ varies with $z$ along the center axis,
exhibiting the maximum value at $z=0$ which thus provides a
quantum point contact (constriction) within the channel of the nanocolumn.
This qualitative finding is a consequence of surface pinning in combination with the doping profile
on a nanometer scale. It is a unique feature of the considered nanocolumn system,
not found in bulk device structures and does not require any built-in heterostructure.

Four relevant length scales exist for this system (see also Fig.~\ref{fig:column}):
The column diameter $W$,
the intrinsic layer thickness $D$,
the homogenous Schottky depletion length $d_0$ as defined in Eq.~(\ref{eq:d0}),
and the ``pinning de Broglie wavelength''
\begin{equation}
\label{eq:lambda0}
\lambda_0:=\sqrt{\frac{2\pi^2\hbar^2}{m^*\phi_0}},
\end{equation}
for an effective electron mass $m^*$ and the given $\phi_0$.
For realistic system parameters $D\lesssim d_0<W/2$,
an approximate expression for the potential profile in the vicinity of the saddle-point can be given:
Within the $x,y$-plane at $z=0$, we can write (see also Fig.~\ref{fig:harmpot}(a))
\begin{equation}
\label{eq:potxy}
E_C(x,y,z=0)=\frac{1}{2}m^*\omega_0^2(x^2+y^2)+\phi_m
\end{equation}
with the effective mass $m^*$ and the the harmonic oscillator frequency $\omega_0$.
From the pinning boundary conditions at $\sqrt{x^2+y^2}=W/2$ we readily obtain
\begin{equation}
\label{eq:a}
m^*\omega_0^2=\frac{8(\phi_0-\phi_m)}{W^2}.
\end{equation}
Furthermore, along the $z$-axis at $x=y=0$, we can approximate (see also Fig.~\ref{fig:harmpot}(b))
\begin{equation}
\label{eq:potz}
E_C(x=0,y=0,z)=\phi_m-\frac{1}{2}bz^2
\end{equation}
with a positive constant $b$. The width of this barrier is determined by a length scale $D_0\sim D$
and this yields
\begin{equation}
\label{eq:b}
b=\frac{2\phi_m}{D_0^2}.
\end{equation}
(The proportionality factor between $D_0$ and $D$ is of the order of 1.
In general, it may depend on $D/d_0$ and $D/W$
and can be determined numerically, see Sec.~\ref{sec:3}.)
From the condition $\Delta E_C=0$ we finally obtain for the barrier height
\begin{equation}
\label{eq:barr}
\phi_m=\phi_0\cdot\frac{1}{1+\frac{W^2}{8D_0^2}}
\end{equation}
and the quantization energy within the saddle-point
\begin{equation}
\label{eq:quant}
\hbar\omega_0=
\sqrt{\frac{8\hbar^2\phi_0}{m^*W^2}}
\cdot
\frac{1}{\sqrt{1+\frac{8D_0^2}{W^2}}}.
\end{equation}
This result can be recast as
\begin{equation}
\label{eq:quant2}
\hbar\omega_0=
\phi_0\cdot\frac{2}{\pi}\cdot\frac{\lambda_0}{W}\cdot
\frac{1}{\sqrt{1+\frac{8D_0^2}{W^2}}}.
\end{equation}

In the next section, these findings (Eqs.~(\ref{eq:barr}),(\ref{eq:quant})) will now be verified with the help of
numerical simulations for a realistic nanocolumn example.

\section{Realistic example}
\label{sec:3}
In order to visualize the results of the previous section, we now consider a realistic example.
We choose GaAs as the semiconductor material with an effective mass of $m^*=0.067 m_e$,
$\epsilon_r=13.1$,
a nanocolumn diameter of $W=100$nm,
a doping concentration of $N_D=1\times 10^{18}$cm$^{-3}$,
an intrinsic layer thickness of $D=10$nm,
and a Fermi-level pinning of $\phi_0=0.7$eV.
The system is assumed to be in an equilibrium state at room temperature.

\begin{figure}
\resizebox{0.45\textwidth}{!}{
  \includegraphics{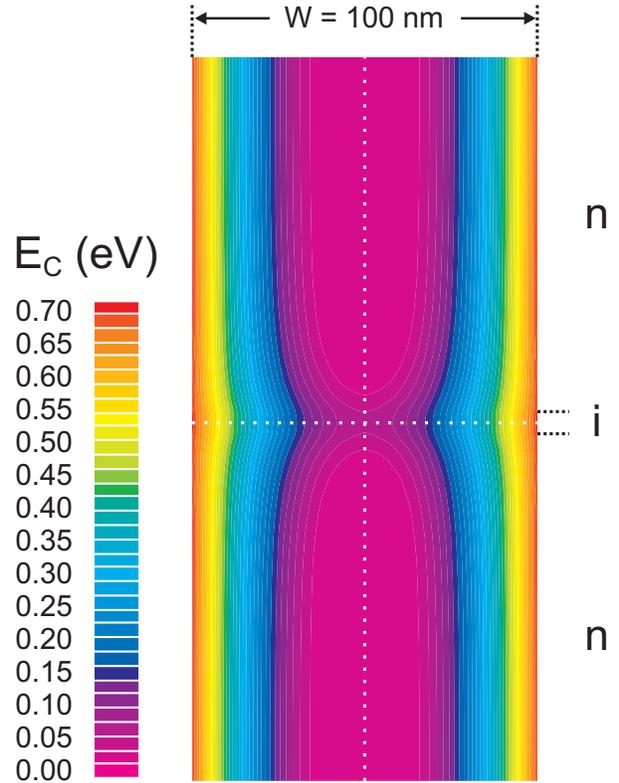}
}
\vspace{2mm}
\caption{
Simulated energy profile $E_C$ within a cylindrical nanocolumn with a diamter of $W=100$nm
with an embedded intrinsic layer of $D=10$nm.
The map shows a cut through the center axis ($z$) of the structure.
The position of the saddle-point is indicated by the two intersecting dotted lines.
}
\label{fig:simmap}
\end{figure}

\begin{figure}
\resizebox{0.45\textwidth}{!}{
  \includegraphics{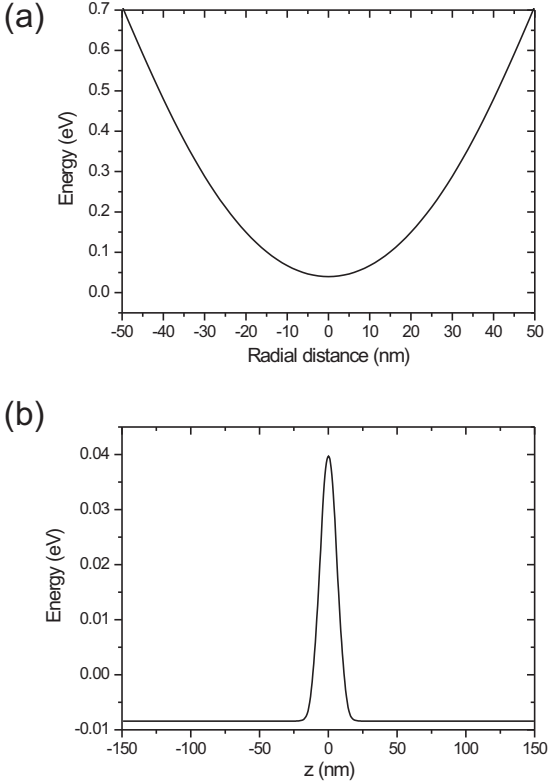}
}
\caption{
(a) Simulated energy profile $E_C$ within the $x,y$-plane (radial distance) at $z=0$.
(b) Simulated energy profile $E_C$ along the center axis $z$ at $x=y=0$.
Note that $E_C(z\to\pm\infty)$ drops below the Fermi-energy $E_F\equiv 0$
due to the assumed doping concentration of $N_D=1\times 10^{18}$cm$^{-3}$.
}
\label{fig:simpot}
\end{figure}

By use of a numerical 3D Poisson-solver \cite{Ref_Silvaco} for cylindrical symmetry,
we have calculated the potential profile within the nanocolumn structure.
Fig.~\ref{fig:simmap} visualizes the results as a color map.
As an important feature of the system, one can clearly identify the
formation of a saddle-point as predicted in Sec.~\ref{sec:2}.
Fig.~\ref{fig:simpot}(a) and (b) furthermore show the details of the potential energy
within the $x,y$-plane at $z=0$ and along the center axis at $x=y=0$, respectively.
(Compare also with Fig.~\ref{fig:harmpot}.)
In turn, a barrier height of $\phi_m=40$meV can be extracted from the simulation.
Furthermore, for the curvature of the $x,y$-potential within the saddle-point we get
$543$eV$\mu$m$^{-2}$.

In order to compare these values with the expressions of Sec.~\ref{sec:2}, we invert
Eq.~(\ref{eq:barr}) and obtain
\begin{equation}
\label{eq:d00}
D_0=W\cdot\frac{1}{\sqrt{8\left(\frac{\phi_0}{\phi_m}-1\right)}}.
\end{equation}
Inserting the simulated value for $\phi_m$, we thus get $D_0=8.7$nm which is in good agreement with $D=10$nm.
Furthermore, employing Eq.~(\ref{eq:a}), we obtain a curvature of $m^*\omega_0^2=528$eV$\mu$m$^{-2}$,
which stands in excellent agreement with the result extracted from the simulation above.
As for the depletion length, we obtain $d_0=32$nm from Eq.~(\ref{eq:d0}), which fits very well to the
results in Fig.~\ref{fig:simmap} for $z\to\pm\infty$.
The analytical expressions given in Sec.~\ref{sec:2} thus represent good approximations
for the estimation of the quantum point contact parameters.

Finally, the lateral quantization energy within the quantum point contact given by Eq.~(\ref{eq:quant})
reads as $\hbar\omega_0=25$meV (the same follows from Eq.~(\ref{eq:quant2}) with the pinning wavelength
$\lambda_0=5.7$nm provided by Eq.~(\ref{eq:lambda0})).
Compared to the energy scale $k_BT=26$meV at room temperature, we thus can expect
a significant impact on the electronic transport properties of a nanodevice based on
such a nanocolumn structure at application-relevant temperatures \cite{Ref_NL5_2470}.


%
%
%

\section{Summary}
\label{sec:summary}
We have shown how nanoscale doping profiles inside a surface-depleted nanocolumn
give rise to the formation of a saddle-point in the potential profile.
The obtained lateral confinement varies within the channel direction,
resulting in an embedded quantum point contact, which can be the origin of novel
collimation transport effects in such nanocolumn structures.

The considered quantum confinement effect is solely due to the formation of
a saddle-point in the potential as a special solution of the 3D Poisson equation
for a nanoscale intrinsic layer in combination with surface pinning boundary conditions,
without the need for any kind of heterostructure.
The presented results can be utilized for a new kind of ``saddle-point potential engineering''
to improve the transport characteristics of quasi-1D nanodevices.

%
%

\end{document}